\documentclass[onecolumn,aps,showpacs,prb,preprint,superscriptaddress]{revtex4}
\usepackage{tipa}
\usepackage{bbding}
\usepackage{txfonts}
\usepackage{amssymb}
\usepackage{graphicx}
\usepackage{CJK}

\begin{document}


\title{{\itshape{Ab initio}} simulations of the kinetic properties of the hydrogen monomer on graphene}

\author{Liang Feng Huang}\affiliation{Key
Laboratory of Materials Physics, Institute of Solid State Physics,
Chinese Academy of Sciences, Hefei 230031, China}
\author{Mei Yan Ni}\affiliation{Key Laboratory of
Materials Physics, Institute of Solid State Physics, Chinese Academy
of Sciences, Hefei 230031, China}\affiliation{School of Electronic
Science and Applied Physics, Hefei University of Technology, Hefei
230009, China}
\author{Xiao Hong Zheng}
\author{Wang Huai Zhou}
\author{Yong Gang Li}
\author{Zhi Zeng}\thanks{Email: zzeng@theory.issp.ac.cn}\affiliation{Key Laboratory of Materials Physics, Institute of
Solid State Physics, Chinese Academy of Sciences, Hefei 230031,
China}

\begin{abstract}
The understanding of the kinetic properties of hydrogen (isotopes)
adatoms on graphene is important in many fields. The kinetic
properties of hydrogen-isotope (H, D and T) monomers were simulated
using a composite method consisting of density functional theory,
density functional perturbation theory and harmonic transition state
theory. The kinetic changes of the magnetic property and the
aromatic $\pi$ bond of the hydrogenated graphene during the
desorption and diffusion of the hydrogen monomer was discussed. The
vibrational zero-point energy corrections in the activation energies
were found to be significant, ranging from 0.072 to 0.205 eV. The
results obtained from quantum-mechanically modified harmonic
transition state theory were compared with the ones obtained from
classical-limit harmonic transition state theory over a wide
temperature range. The phonon spectra of hydrogenated graphene were
used to closely explain the (reversed) isotope effects in the
prefactor, activation energy and jump frequency of the hydrogen
monomer. The kinetic properties of the hydrogen-isotope monomers
were simulated under conditions of annealing for 10 minutes and of
heating at a constant rate (1.0 K/s). The isotope effect was
observed; that is, a hydrogen monomer of lower mass is desorbed and
diffuses more easily (with lower activation energies). The results
presented herein are very similar to other reported experimental
observations. This study of the kinetic properties of the hydrogen
monomer and many other involved implicit mechanisms provides a
better understanding of the interaction between hydrogen and
graphene.
\end{abstract}


\maketitle

\section{INTRODUCTION}
\par The fact that hydrogen interacts with graphite surface in outer
space\cite{coey,hornekaer96,hornekaer97,ferro368} and in fusion
devices\cite{ferro368,morris} requires a clear understanding of the
kinetic properties of hydrogen adatoms on graphite surfaces for
astronomy and the nuclear industry. Such properties are also very
important for the realization of both graphite-based hydrogen
storage\cite{anthony} and graphene-based
electronics.\cite{shytov,ryu,elias,sofo75,balog131,bostwick103,guisinger9,luo3,lebegue79}

\par The desorption and diffusion of hydrogen adatoms on
graphite surfaces are two important kinetic processes whose
properties depend sensitively on the interaction between the adatom
and the graphite surface. Consequently, investigation of the
desorption and diffusion of hydrogen adatoms on graphite surfaces
can further our understanding of the mechanisms involved in these
interactions. Some researches have been carried out on the kinetic
properties of hydrogen adatoms on graphite
surfaces.\cite{hornekaer96,hornekaer97,zecho42,zecho117} The
adsorption of hydrogen monomers on graphene is an essential step in
hydrogenation, and the kinetic properties of this step determine the
outcome of the hydrogenation process. Hornek{\ae}r et. al. found
that a large fraction of the D monomers on a graphite surface will
be desorbed during annealing at room temperature for 10 minutes,
while the D monomers on the graphite surface are diffusionally
immobile at room temperature.\cite{hornekaer97} Such diffusional
immobility is consistent with Baouche's observations in
time-programmed desorption experiments.\cite{baouche125} Some
theoretical studies on the kinetic properties of hydrogen monomers
have also been
reported.\cite{hornekaer97,ferro368,ferro120,casolo130,herrero79,roman45}
The calculated chemisorption energies of hydrogen monomer on
graphene vary around a value of 0.75 eV with a range of about 0.3
eV; this magnitude of variation represents quite a large energetic
uncertainty in kinetics. Furthermore, in kinetic simulations, the
vibrational characteristics of the light-mass hydrogen monomers
should be considered. Thus, a fully theoretical simulation that can
be precisely and directly compared with experimental observations in
both desorption and diffusion is still lacking.

\par In this report, the kinetic properties of various hydrogen-isotope (H,
D and T) monomers were simulated by means of a composite method
consisting of density functional theory (DFT),\cite{kohn140} density
functional perturbation theory (DFPT)\cite{baroni73} and harmonic
transition state theory
(hTST).\cite{eyring3,vineyard3,hanggi62,pollak15} The kinetic change
of the magnetic property of the hydrogenated graphene during the
desorption and diffusion of the hydrogen monomer was discussed. The
vibrational contribution, including the zero-point energy correction
in the activation energy, was considered with hTST. The kinetic
properties of various hydrogen-isotope monomers were simulated under
conditions of annealing for 10 minutes and of heating at a constant
rate (1.0 K/s).

\section{METHODOLOGY}

\par The over-barrier jump frequency between two local
minimum states (initial and final states or reactant and product
states) can be expressed in the Arrhenius form
as\cite{hanggi62,toyoura78}
\begin{eqnarray}{\label{arrhenius}}
v=v^*\exp{(-\frac{E_{ac}}{k_BT})}
\end{eqnarray}
where $v^*$ is the exponential prefactor and $E_{ac}$ is the
activation energy that is required for the reaction to occur. The
activation energy is defined herein to be the vibrational zero-point
energy corrected potential barrier, which is expressed as
\begin{eqnarray}{\label{activation}}
E_{ac} & = & {\Delta}E_{p}+\frac{1}{2}\sum_{i=1}^{3N-1}\hbar{\omega}_i^S-\frac{1}{2}\sum_{i=1}^{3N}\hbar{\omega}_i^I \nonumber\\
       & = & {\Delta}E_{p}-{\Delta}F_{vib}(0)
\end{eqnarray}
where ${\Delta}E_p$ is the potential barrier in the reaction path,
which can be obtained from DFT calculations, and ${\omega}^I_i$ and
${\omega}^S_i$ are the vibrational frequencies of the $i$th mode in
the initial and saddle-point states, respectively. Correction
${\Delta}F_{vib}(0)$ comes from the vibrational zero-point energy
difference between the initial and saddle-point states. The total
vibrational degrees of freedom are 3N; an imaginary vibrational mode
along the migration coordinate in the saddle-point state is
excluded; thus, 3N-1 vibrational modes are considered for the
saddle-point state. From quantum-mechanically modified hTST,
prefactor is expressed as\cite{eyring3,toyoura78,sundell76}
\begin{eqnarray}{\label{qmTST}}
v^*_{qm}=\frac{k_BT}{h}\frac{\prod\limits_{i=1}^{3N}[1-\exp{(-\frac{\hbar\omega_i^I}{k_BT})}]}{\prod\limits_{i=1}^{3N-1}[1-\exp{(-\frac{\hbar\omega_i^S}{k_BT})}]}
=\frac{k_BT}{h}\frac{\prod\limits_{i=1}^{3N-1}\exp{(\frac{\hbar\omega_i^S}{k_BT})}\bar
n_T(\omega_i^S)}{\prod\limits_{i=1}^{3N}\exp{(\frac{\hbar\omega_i^I}{k_BT})}\bar
n_T(\omega_i^I)}
\end{eqnarray}
where $\bar n_T(\omega_i)$ is the bosonic phonon occupation number
of the $i$th vibrational mode. When the temperature approaches
infinity, the classical limit of the prefactor is expressed as
\begin{eqnarray}{\label{clTST}}
v^*_{cl}=\frac{1}{2\pi}\frac{\prod\limits_{i=1}^{3N}\omega_i^I}{\prod\limits_{i=1}^{3N-1}\omega_i^S}
\end{eqnarray}
This classical-limit form is also the Vineyard's
form,\cite{vineyard3} in which all of the vibrational modes are
assumed to be completely thermo-activated. From hTST as described
above, the mass-dependent vibrational frequencies are responsible
for the difference in kinetic properties of H, D and T monomers as
shown below in this report.

\par The first-order rate equation for desorption can be
expressed as\cite{baouche125,hanggi62}
\begin{eqnarray}{\label{first_order}}
\frac{dn(t)}{dt}=-{v_{des}(T)}n(t)
\end{eqnarray}
where $v_{des}$ is the desorption frequency; $n(t)$ is the residual
number of hydrogen monomers on graphene at the $t$ moment, and $t=0$
represents the starting moment for initializing the kinetic
movement.
\par In the annealing process, the temperature is kept constant ($T=T_0$);
in that case, the variation of the residual number with respect to
time is expressed as
\begin{eqnarray}{\label{n_anneal}}
n(t)=n(0)\exp{[-v_{des}(T_0)t]}
\end{eqnarray}
The diffusion can be described by the mean square displacement
($<|{\bf r}(t)-{\bf r}(0)|^2>$) of a monomer parallel to graphene.
From Fick's second law, we have
\begin{eqnarray}
<|{\bf{r}}(t)-{\bf{r}}(0)|^2>=2dD_{ad}(T)t
\end{eqnarray}
where $d$ is the dimensionality of the diffusion of a hydrogen
monomer on graphene (taken to be 2 here) and $D_{ad}$ is the
diffusion coefficient, which is temperature-dependent and is
expressed as\cite{toyoura78}
\begin{eqnarray}
D_{ad}=\frac{1}{2d}{\Gamma}a^2
\end{eqnarray}
where $\Gamma$ is the total jump frequency of the monomer and $a$ is
the jump length. For the diffusion of a monomer on graphene,
$\Gamma$ is taken here to be $3v_{diff}$ ($v_{diff}$ is the
diffusion frequency), which is combined with the fact that there are
three equivalent final sites to which a monomer at an initial site
can diffuse, and $a$ is taken to be the optimized C--C bond length
of 1.426 \AA{}. The diffusion radius of the monomer is defined as
the square root of the mean square displacement
\begin{eqnarray}{\label{diffusion radius}}
r_{diff}=\sqrt{2dD_{ad}(T)t}
\end{eqnarray}
which can be used directly to determine whether the hydrogen monomer
on graphene is diffusionally mobile.
\par In the heating process, if the temperature increases at a constant rate $\alpha$
($T={\alpha}t$), the variation of the residual number with respect
to time is expressed as
\begin{eqnarray}{\label{n_heating}}
n(t)=n(0)\exp{[-\int_0^tv_{des}({\alpha}t)dt]}
\end{eqnarray}
where $\alpha$ is always taken to be 1.0
K/s.\cite{hornekaer96,zecho42,zecho117} The relative desorption rate
is therefore defined by the ratio of the desorption rate to the
initial monomer number
\begin{eqnarray}{\label{desorption rate}}
R_{des}(T,t)=v_{des}(T)n(t)/n(0)
\end{eqnarray}
which is both time- and temperature-dependent.

\par In this report, a monolayer of graphene served as a
structural model of a graphite surface. This was feasible because
the effect of the weak van de Waals interaction between neighboring
graphene layers on the chemisorption properties of hydrogen monomer
can be neglected; in our DFT tests using local density approximation
(LDA) and generalized gradient approximation (GGA), the
bilayer-graphene structural model yields potential barriers with a
difference less than 8 meV with respect to the monolayer structural
model. The lattice constant of graphene has been found very close to
graphite (within 0.05\%) at 200 -- 500 K,\cite{mounet71} which is
the most important temperature range for the kinetic movements of
the hydrogen monomer as shown below. A hydrogen monomer on a
$5\times5$ periodic supercell of graphene (50 C atoms) with a 10
\AA{} vacuum along the direction perpendicular to the surface can be
considered as an isolated monomer; our tests show that the
interaction between two monomers in two such neighboring supercells
can be neglected. This is consistent both with the findings of
Shytov\cite{shytov} based on a theory of electron-mediated
interaction and with those of Boukhvalov\cite{boukhvalov21} based on
DFT calculations. The DFT and DFPT calculations were carried out
using the Quantum Espresso code package,\cite{giannozzi21} in which
the ultrasoft\cite{vanderbilt41} spin-polarized PBE\cite{perdew77}
pseudopotentials were used to describe the electronic exchange and
correlation energy. The wave function and the charge density were
expanded using energy cutoffs of 35 and 350 Ry. For the calculation
of electronic density of states (DOS), a $6\times6\times1$ uniform
$k$-point grid with the tetrahedron interpolation
scheme\cite{blochl49} was used for the integration of the electronic
states over the first Brillouin zone. And for the other
calculations, a $4\times4\times1$ uniform $k$-point grid with the
Methfessel-Paxton smearing technique\cite{methfessel40} was used,
where the smearing width was taken to be 0.03 Ry. The partial DOS of
hydrogen was obtained using the L\"owdin population analysis. A
force threshold of $10^{-4}$ Ry/bohr was used for the structural
optimization. The reaction paths were described by the minimum
energy paths (MEPs) between two local minimum states that were
calculated using the climbing-image nudged elastic band
method\cite{henkelman113} with 5 images for each reaction path. For
calculating the vibrational frequencies, only the gamma point at the
Brillouin zone center was selected.

\section{RESULTS AND DISCUSSION}
\par
\begin{figure*}[htbp]
\scalebox{0.25}[0.25]{\includegraphics{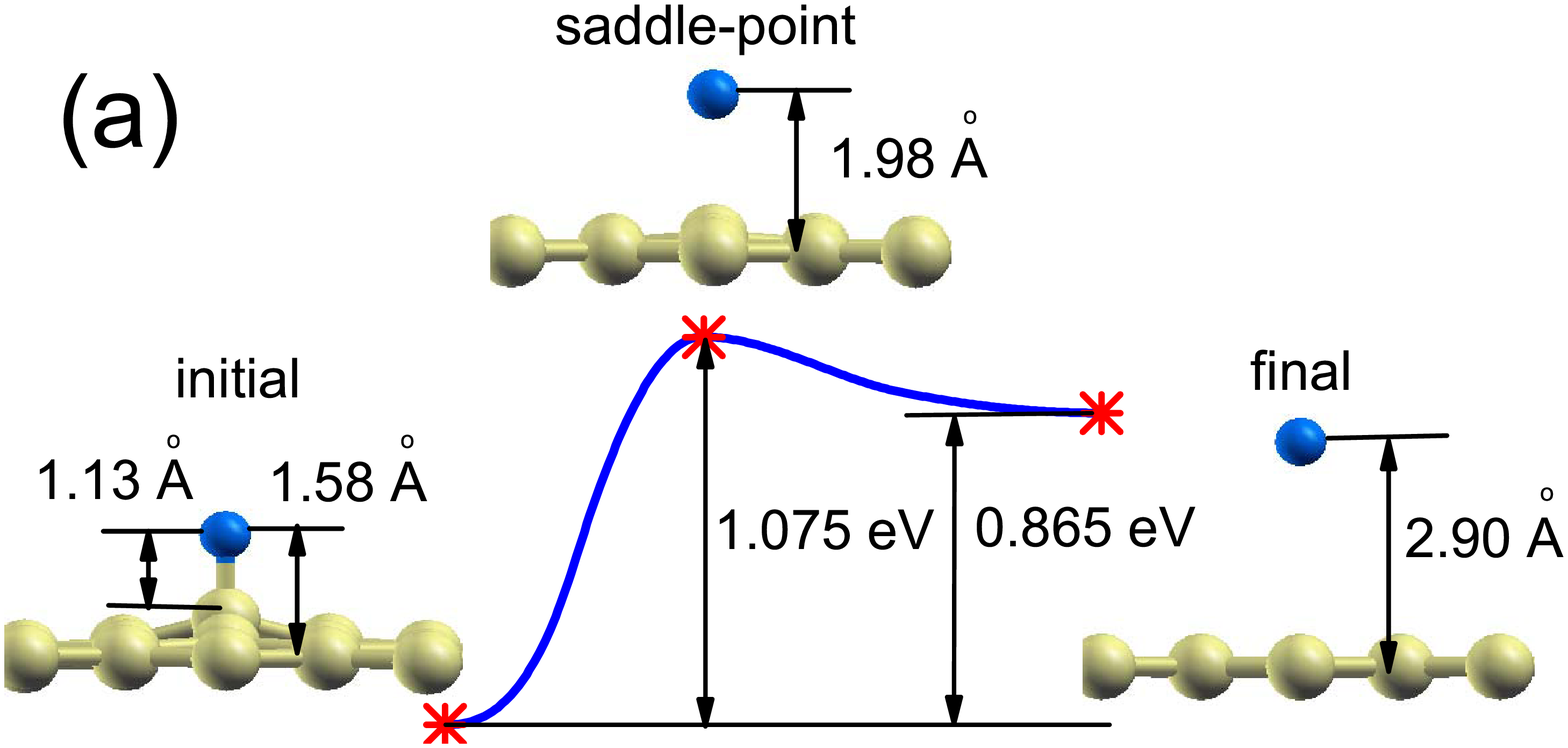}}\\
\scalebox{0.25}[0.25]{\includegraphics{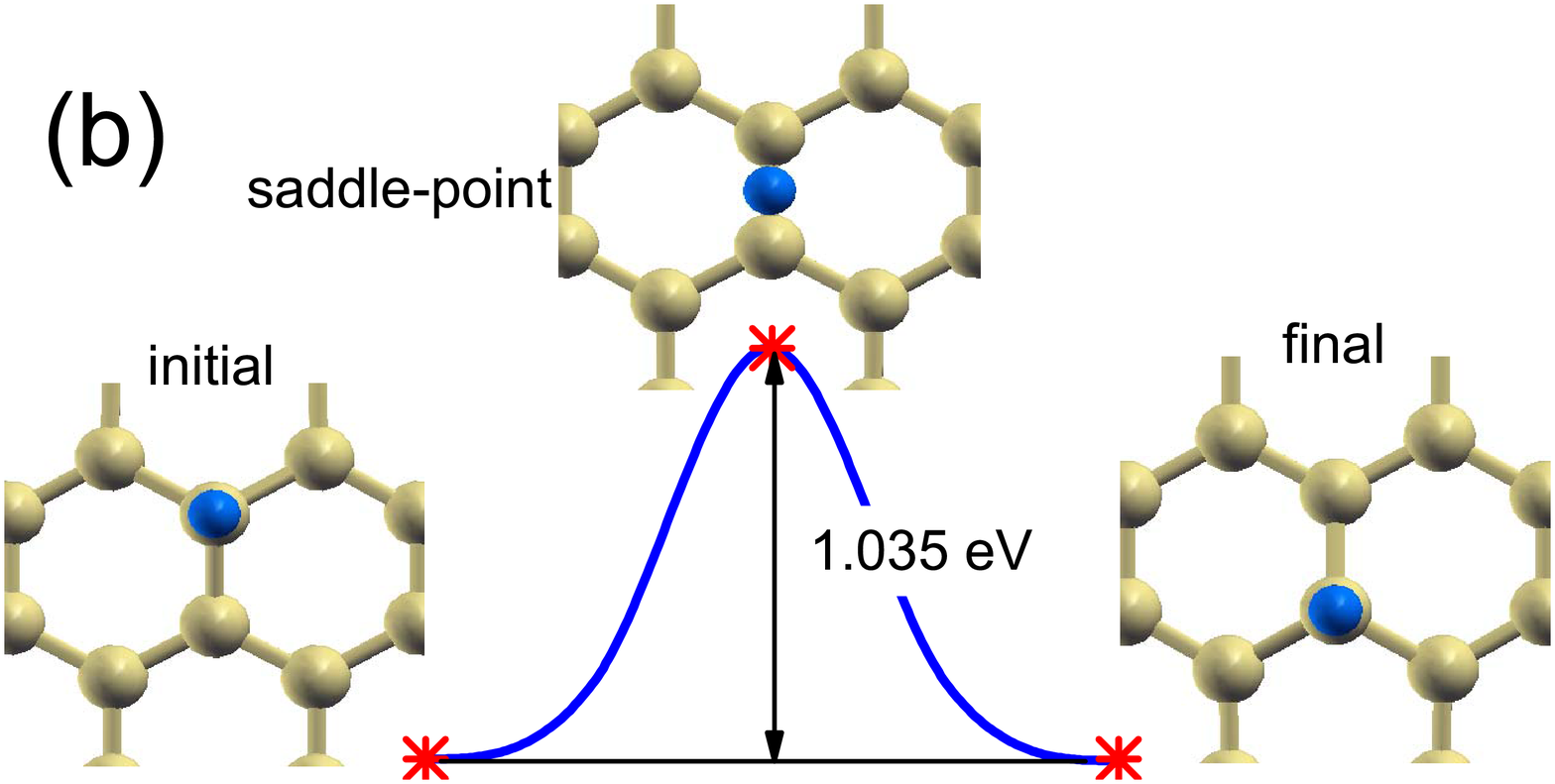}}
\caption{\label{MEPs}(Color online) MEPs for the (a) desorption and
(b) diffusion of a hydrogen monomer on a graphene layer. In the
paths, the initial, saddle-point and final states are labeled with
stars, and their structures are shown at the side of the figure. The
final state in the desorption MEP is fixed to be the physisorption
state. The yellow spheres represent carbon atoms and the smaller
blue spheres represent hydrogen atoms.}
\end{figure*}

 The calculated MEPs for the desorption and diffusion processes of a hydrogen monomer
are shown in Fig. \ref{MEPs}, in which the structures of the initial
(reactant), saddle-point (transition) and final (product) states are
also shown. The final state in the desorption MEP has been fixed to
be the physisorption state. In order to simulate the desorption MEP
from the physisorption state to the fully desorbed state, a vacuum
with a size larger than 10 \AA{} (e.g. 16.6 \AA{} in Ref.
\onlinecite{hornekaer97}) is needed to completely eliminate any weak
interaction between the desorbed hydrogen atom and the graphene
layer in the neighboring supercell. However, as seen from the
methodology part, precise potential surfaces from the chemisorption
state to the saddle-point states are enough for the simulation of
reaction rates in this report. The energy for the fully desorbed
state here is calculated by summing the energies of an isolated
hydrogen atom and an isolated graphene layer. In the desorption
process shown in Fig. \ref{MEPs} (a), the hydrogen monomer moves
away from the graphene layer in the direction perpendicular to the
layer. At the first stage of the desorption process (the
chemisorption state), the hydrogen atom is at the top site with a
height of 1.58 \AA{} bonded to a C atom that protrudes from the
graphene plane with a height of 0.45 \AA{}. The calculated
protrusion height is somewhat method-dependent and values ranging
from 0.26 to 0.49 \AA{} have been obtained by
others.\cite{ferro120,miura93,boukhvalov77,ferro78,kerwin128,denis907,peeters114}
In the desorption saddle-point state, the hydrogen atom is at the
top site with a height of 1.98 \AA{}. The potential barrier was
defined as the energy difference between the saddle-point state and
the initial state; the calculated desorption potential barrier is
1.075 eV. The final (physisorption) state is 0.865 eV higher in
energy than the initial (chemisorption) state; the fully desorbed
state is 0.890 eV higher than the initial state and 0.035 eV higher
than the physisorption state. The small energy difference of 0.035
eV also indicates that very large vacuum is needed if one wants to
obtain the fully desorbed state in the vacuum. This calculated
potential surface along the desorption MEP is very close to
Hornek{\ae}r's theoretical result,\cite{hornekaer97} where the
smoothly varying potential surface for the MEP from the
physisorption state to the fully desorbed state also has been shown.
In addition, if the hydrogen monomer climbs over the desorption
barrier, it will escape from the graphene layer with high velocity
after climbing down the potential barrier. In the diffusion process
shown in Fig. \ref{MEPs}(b), the hydrogen atom moves along a C--C
bond at an average height of about 1.40 \AA{}. In the diffusion
saddle-point state, the hydrogen atom is at the bridge-site with a
height of 1.20\AA{}. The diffusion potential barrier is 1.035 eV,
which is somewhat smaller than the desorption potential barrier.

\par
\begin{figure}[htbp]
\scalebox{0.4}[0.4]{\includegraphics{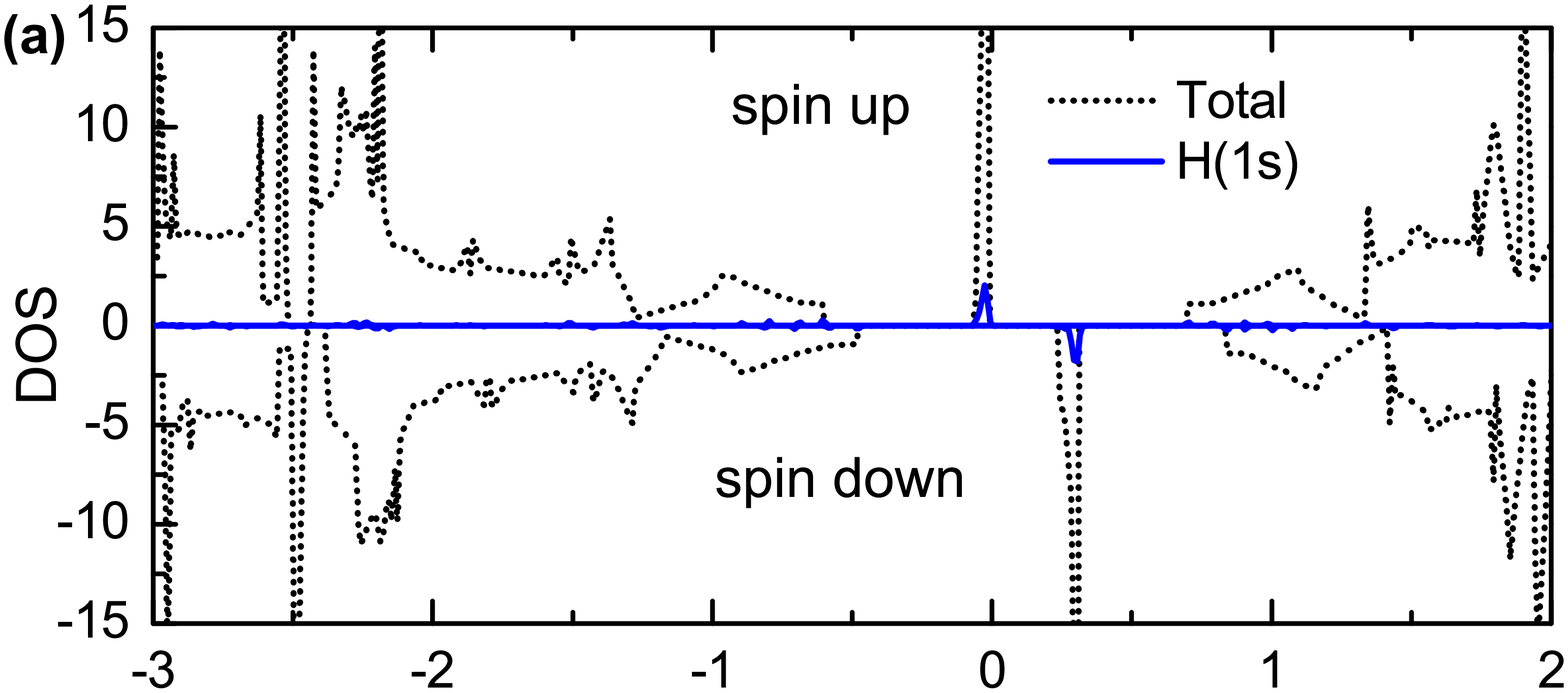}}\\
\scalebox{0.4}[0.4]{\includegraphics{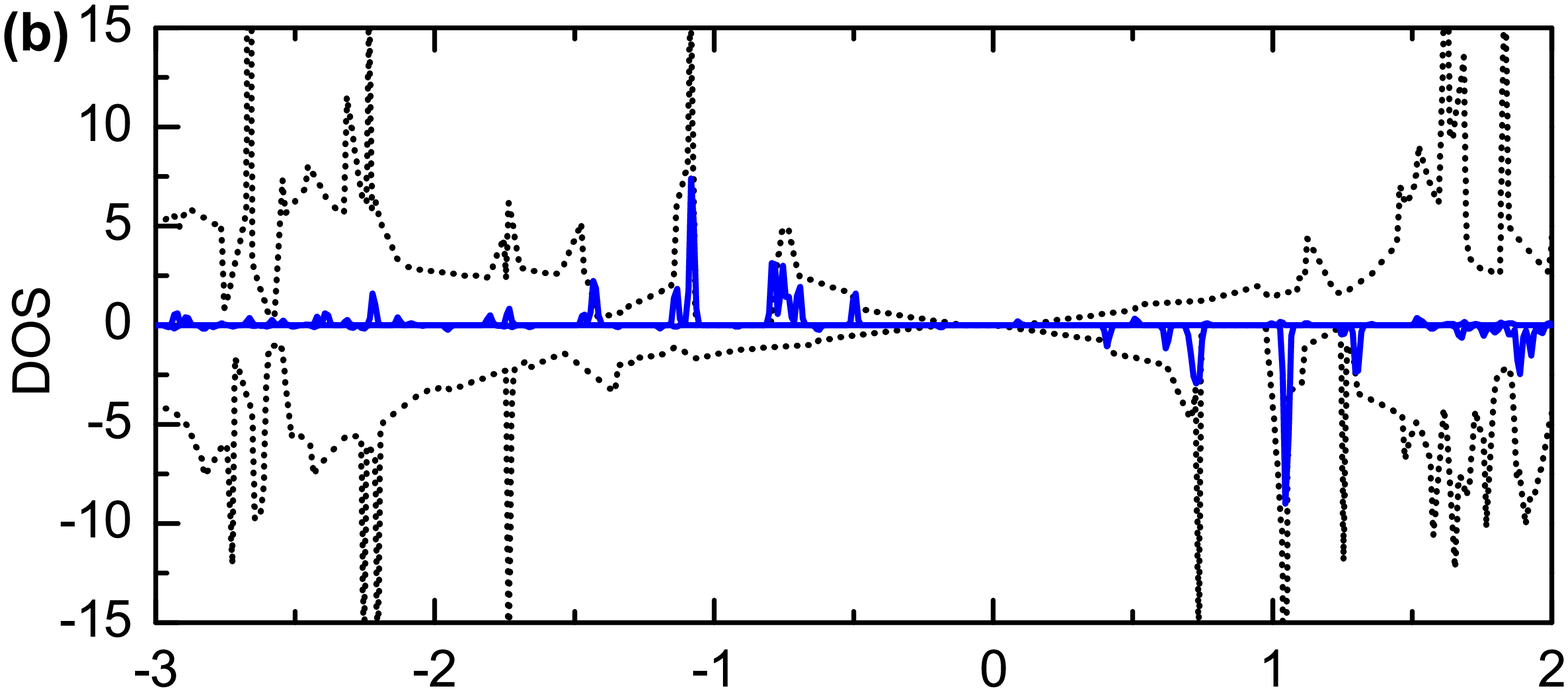}}\\
\scalebox{0.4}[0.4]{\includegraphics{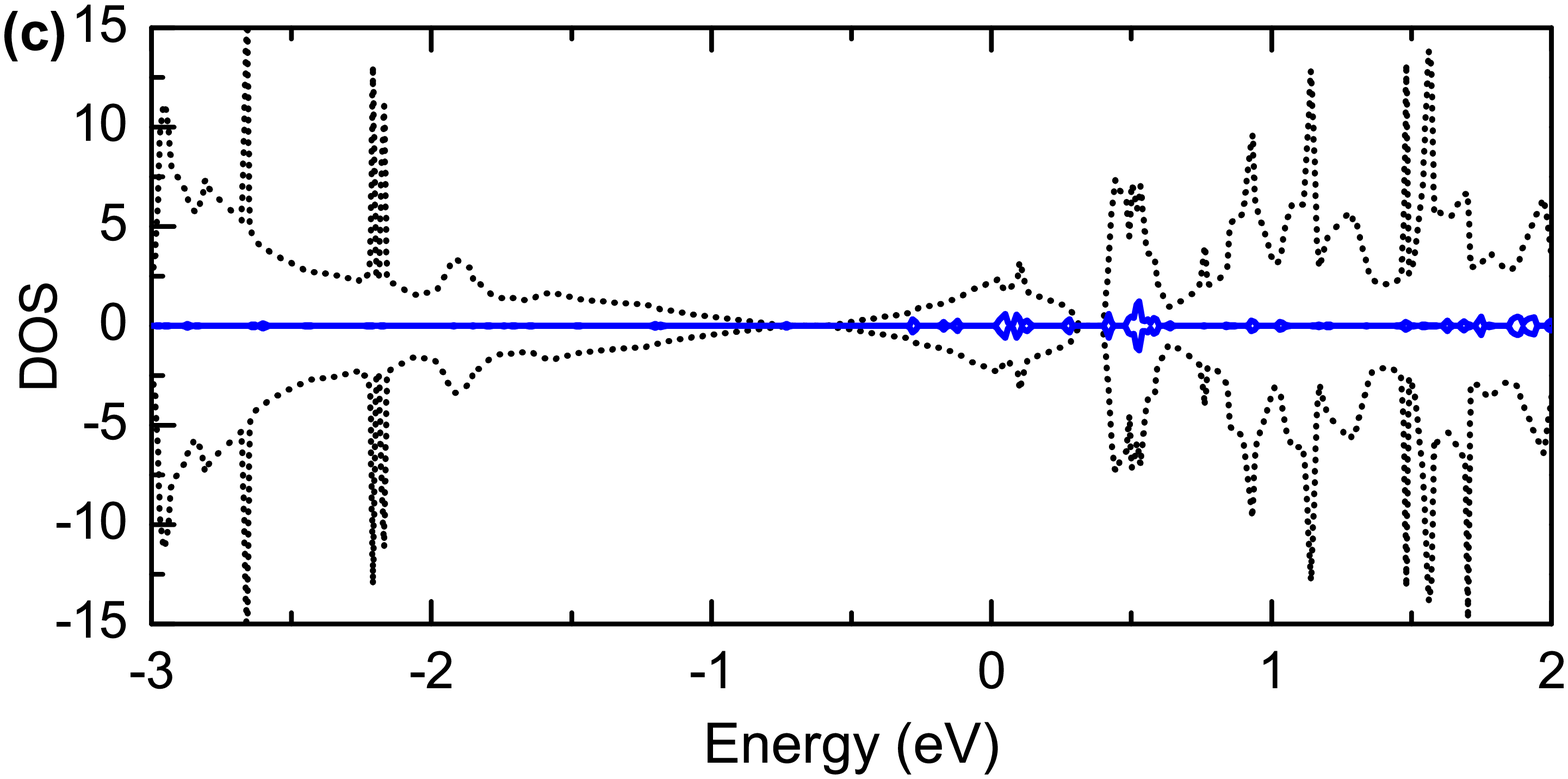}}
\caption{\label{DOS}(Color online) The total electronic DOS and the
partial DOS of the hydrogen atom of the (a) initial state, (b)
desorption saddle-point state and (c) the diffusion saddle-point
state.}
\end{figure}

 The kinetic movement of the hydrogen monomer can result in the
kinetic change of the magnetic property of the hydrogenated
graphene. Fig. \ref{DOS} shows the electronic DOS of the initial,
desorption saddle-point and diffusion saddle-point states, as well
as the respective partial DOS of the hydrogen monomer. The initial
state and the desorption saddle-point state both have a total
magnetic moment of 1.0 $\muup_B$, while the diffusion saddle-point
state is non-magnetic. The DOS of the initial state is consistent
with the theoretical results by others.\cite{casolo130,duplock92}
The adsorption of hydrogen atom breaks an aromatic $\pi$ bond in
graphene, which results in one unsaturated C($p$) orbital. This
unsaturated C($p$) orbital is responsible for the two narrow peaks
(bands) placed around the Fermi level in the total DOS. The energy
gap is 0.245 eV. As seen from the partial DOS of the H($1s$)
orbital, the contribution from the H(1$s$) orbital to these two
bands is small. The unpaired electron occupying the lower band is
responsible for the magnetic moment of 1.0 $\muup_B$. The energies
of the other spin-up bands shift downwards a little with respect to
their spin-down counterparts due to the exchange-correlation
interaction between the electrons in the magnetic C($p$) orbital and
the other electrons. In the desorption saddle-point state, the
partial DOS of the H(1$s$) orbital is much larger than that in the
initial state. This is because the H(1$s$) orbital is much less
hybridized with graphene in the desorption saddle-point state than
that in the initial state, and less electrons transfer from H to
graphene. Thus in the desorption process, the contribution of the
H(1$s$) orbital to the total magnetic moment of 1.0 $\muup_B$ will
increase up to 100\% when the H monomer is fully desorbed. The
spin-up bands do not significantly shift with respect to the
spin-down ones, which indicates the interaction between the magnetic
H atom with graphene is weak. And the linear dispersion of the total
DOS near the Fermi level is the same as that of grpahene, indicating
that the aromatic $\pi$ bond in graphene which is broken in the
chemisorption state has been restored. In the diffusion saddle-point
state, the linear dispersion part of the total DOS is not influenced
by the presence of the hydrogen monomer on the bridge site and the
Fermi level shifts upwards into the conduction bands. This reveals
that the hydrogen monomer does not hybridize with graphene but just
dopes graphene with itinerant electrons. This is the reason why the
diffusion saddle-point state is nonmagnetic. Thus in the diffusion
process, the total magnetic moment decreases from 1.0 $\muup_B$ down
to 0.0 $\muup_B$ when leaving from the chemisorption state to the
diffusion saddle-point state.

\par
\begin{figure}[htbp]
\scalebox{0.6}[0.6]{\includegraphics{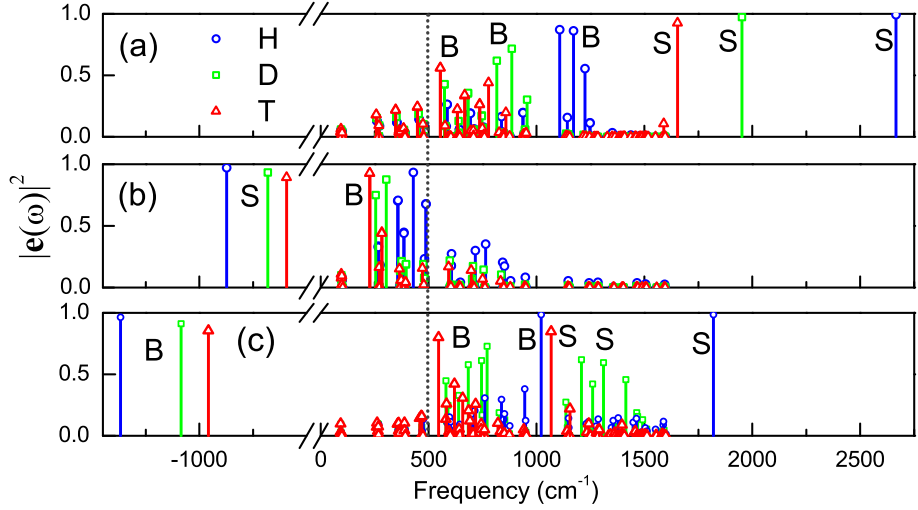}}
\caption{\label{Phonons}(Color online) The calculated spectra of the
$|{\bf{e}}(\omega_i)|^2$ ($i$ is the index of the vibrational mode)
for the hydrogen-isotope monomers in the (a) initial, (b) desorption
saddle-point and (c) diffusion saddle-point states. The imaginary
vibrational modes in the saddle-point states are designated with
negative frequencies. "S" and "B" represent the stretching and
bending modes, respectively. The division of the lower region and
higher region (see text) at 500 cm $^{-1}$ is labeled by a gray
dotted vertical line.}
\end{figure}

 The displacement of the hydrogen monomer in each eigenvector
of the vibrational dynamic matrix was represented by ${\bf
e}(\omega_i)$ ($i$ is the index of the vibrational mode); the
calculated spectra of the $|{\bf e^M}(\omega_i)|^2$ (M=H, D and T)
for the initial, desorption saddle-point and diffusion saddle-point
states are shown in Fig. \ref{Phonons}. The modes with large values
of $|{\bf e}(\omega_i)|^2$ are localized stretching (S) and bending
(B) modes. The isotope effect in the phonon spectra is obvious in
that the vibrational frequency decreases with increasing monomer
mass. The stretching modes of H at 2664 cm$^{-1}$ and of D at 1951
cm$^{-1}$ in the initial state are very close to the experimental
measurements of 2650--2680 cm$^{-1}$ and 1950--1955
cm$^{-1}$.\cite{zecho117,allouche123} In the desorption saddle-point
state, the stretching mode of each isotope monomer becomes
imaginary, while in the diffusion saddle-point state, one bending
mode of each isotope monomer becomes imaginary. The real stretching
and bending modes in the diffusion saddle-point state have larger
vibrational frequencies than the real bending modes in the
desorption saddle-point state; this will result in a difference
between the exponential prefactor of desorption and that of
diffusion, as shown below. For light-mass atoms such as  hydrogen
isotopes, the vibrational zero-point energy corrections
(${\Delta}F_{vib}(0)$) are very important to the precise estimation
of activation energies. The calculated ${\Delta}F_{vib}(0)$s for H,
D and T monomers are listed in Tab. \ref{ZPE}. The values, which
range from 0.072 eV to 0.205 eV, indeed cannot be neglected if the
kinetic properties of hydrogen isotope monomers are to be
understood. The isotope effect in the ${\Delta}F_{vib}(0)$s, which
exhibits as that ${\Delta}F_{vib}(0)$ decreases with increasing
monomer mass, is a result of the isotope effect in the phonon
spectra. Both of these isotope effects will result in the presence
of an isotope effect in the desorption and diffusion frequencies of
hydrogen monomers, as shown in the following. In addition, the
${\Delta}F_{vib}(0)$ correction for the desorption potential barrier
is larger than that for the diffusion potential barrier due to the
disappearance of the highest-frequency real stretching modes and the
significant lowering of the bending modes in the desorption
saddle-point state with respect to those modes in the initial state.
However, in the diffusion saddle-point state, only one
lower-frequency real bending mode disappears and the real stretching
and bending modes are less affected. The resonant in-band modes
(modes with small values of $|{\bf e}(\omega_i)|^2$) in these three
states are slightly shifted and are nearly independent of the
isotopic character of the adatom. The mode with a frequency around
630 cm$^{-1}$ observed in experiments is an example of the resonant
in-band modes.\cite{zecho117,allouche123}

\par
\begin{figure}[htbp]
\scalebox{0.3}[0.3]{\includegraphics{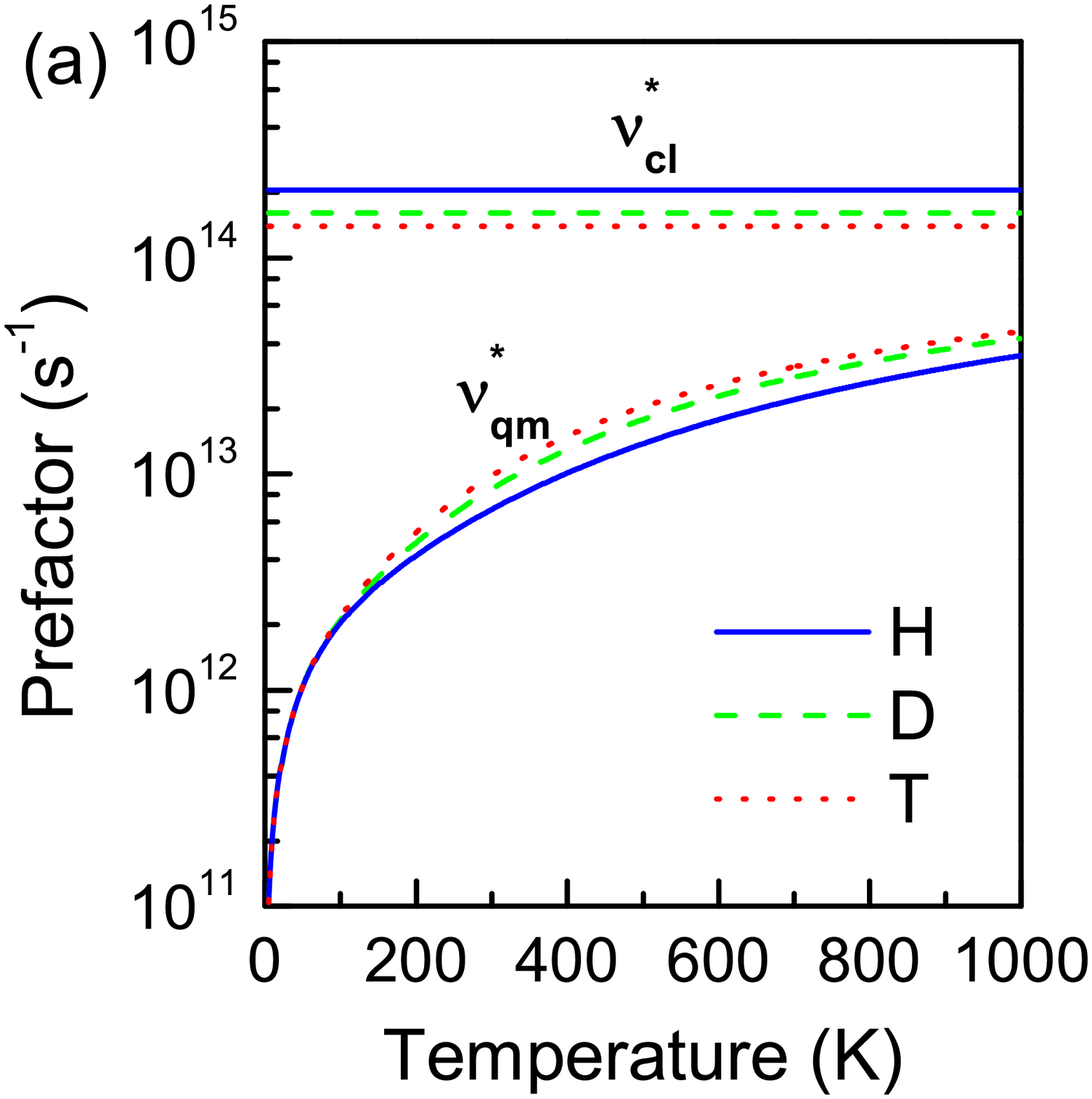}}
\scalebox{0.3}[0.3]{\includegraphics{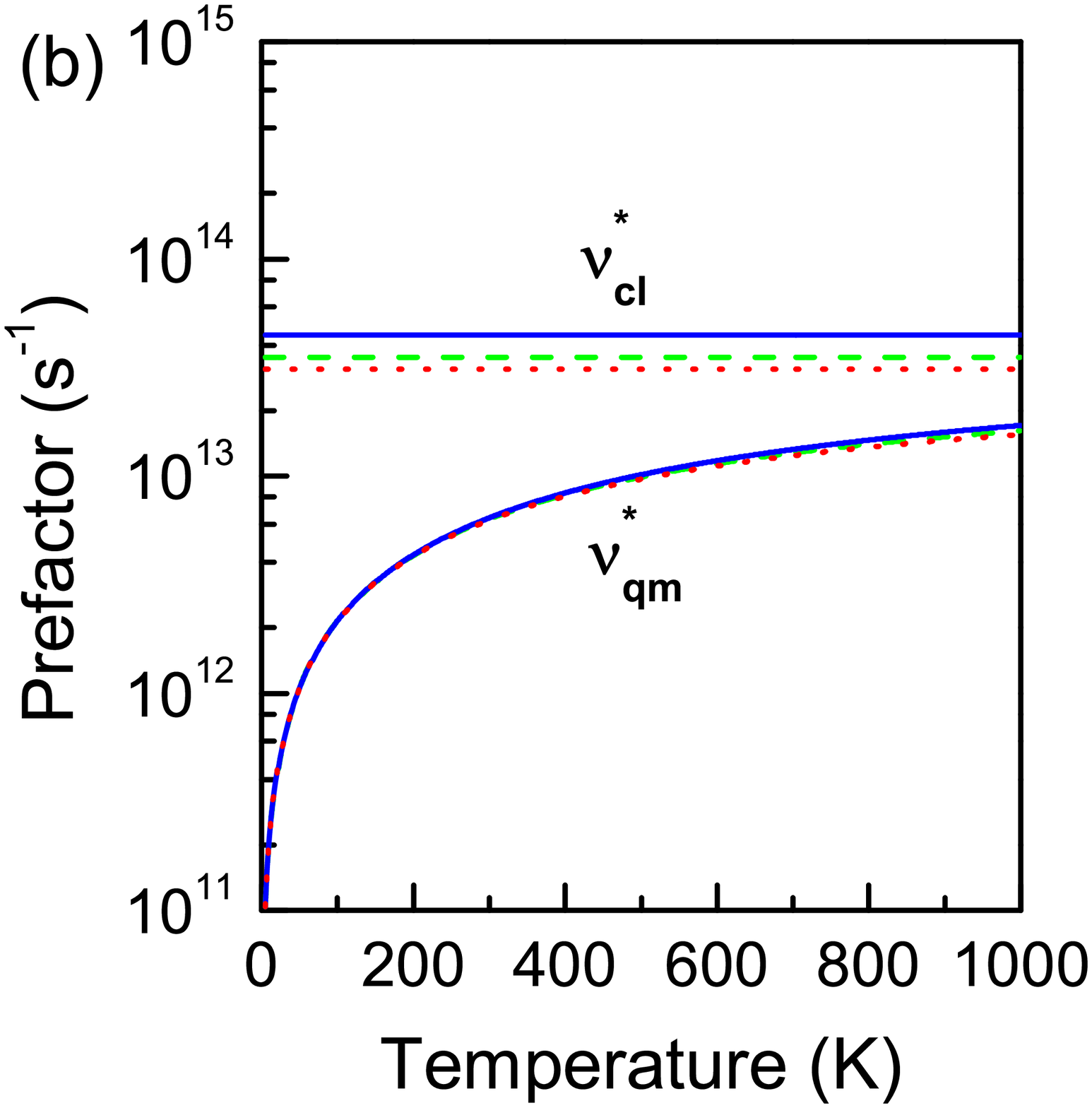}}
\caption{\label{Prefactor}(Color online) Variation of the
exponential prefactors with respect to temperature for the (a)
desorption and (b) diffusion processes of the hydrogen-isotope
monomers on graphene. The values from the quantum-mechanically
modified hTST ($v_{qm}^*$) are compared with their classical-limit
values ($v_{cl}^*$).}
\end{figure}

 The exponential prefactors for H, D and T monomers calculated from the quantum-mechanically
modified hTST ($v_{qm}^*$) and its classical limit ($v_{cl}^*$, the
Vineyard form\cite{vineyard3}) are shown in Fig. \ref{Prefactor}.
The $v_{cl}^*$s are much larger than the corresponding $v_{qm}^*$s
in the plotted temperature range. Because differences in hydrogen
monomer prefactors are mainly determined by the localized stretching
and bending modes, and the contributions from the high-frequency
localized modes are fully considered in the classical-limit hTST
(Equ. \ref{clTST}), while they are much less considered in the
quantum-mechanically modified hTST due to the weight factors of the
bosonic phonon occupation numbers (Equ. \ref{qmTST}). The
quantum-mechanically modified and the classical-limit prefactors
should be equal at very high temperatures. In the classical-limit
hTST, all vibrational modes are regarded as completely
thermo-activated. In the $v_{cl}^*$s, the isotope effect is obvious:
the prefactor decreases with increasing monomer mass, which is due
to the isotope effect in the phonon spectra in Fig. \ref{Phonons},
as discussed in the previous paragraph. However, among the
quantum-mechanically modified desorption prefactors, the
relationship $v_{qm}^*$(T) $>$ $v_{qm}^*$(D) $>$ $v_{qm}^*$(H) holds
and the isotope effect is reversed with respect to the isotope
effect in the $v_{cl}^*$s. The values of the quantum-mechanically
modified diffusion prefactors for H, D and T monomers are nearly the
same over a very wide temperature range (0-1000K). This can be
understood from Equ. \ref{qmTST} and the phonon spectra in Fig.
\ref{Phonons}. For convenience, the phonon spectra are divided into
lower and higher regions below and above 500 cm$^{-1}$,
respectively. In the initial state, all the localized modes are in
the higher region. The real localized bending modes in the
desorption saddle-point state are significantly lowered such that
they now fall within the lower region, while the real localized
modes in the diffusion saddle-point state remain in the higher
region. In the plotted temperature range, the $v_{qm}^*$s are mainly
determined by the vibrational modes in the lower region, due to the
large bosonic phonon occupation numbers of these modes. In the
phonon spectra of the initial and diffusion saddle-point state, the
lower region contains only the resonant in-band modes, which are
nearly independent of the isotopic character of the monomer,
therefore, the $v_{qm}^*$s for H, D and T monomers in diffusion are
nearly the same. However, in the phonon spectra of the desorption
saddle-point state, the low-lying localized bending modes are
significantly occupied by phonons at temperatures above 100 K; in
this state, the localized modes for the T monomer are the lowest and
should be occupied by most phonons and most readily
thermo-activated, followed by the modes for D and H monomers in that
order. This order of thermo-activation leads to the relative
magnitude order of the $v_{qm}^*$s for H, D and T monomers at
temperatures above 100 K; however, this magnitude order will be
progressively altered as the temperature increased to high values
(not shown), finally achieving the same order as $v_{cl}^*$s.

\par
\begin{figure}[htbp]
\scalebox{0.3}[0.3]{\includegraphics{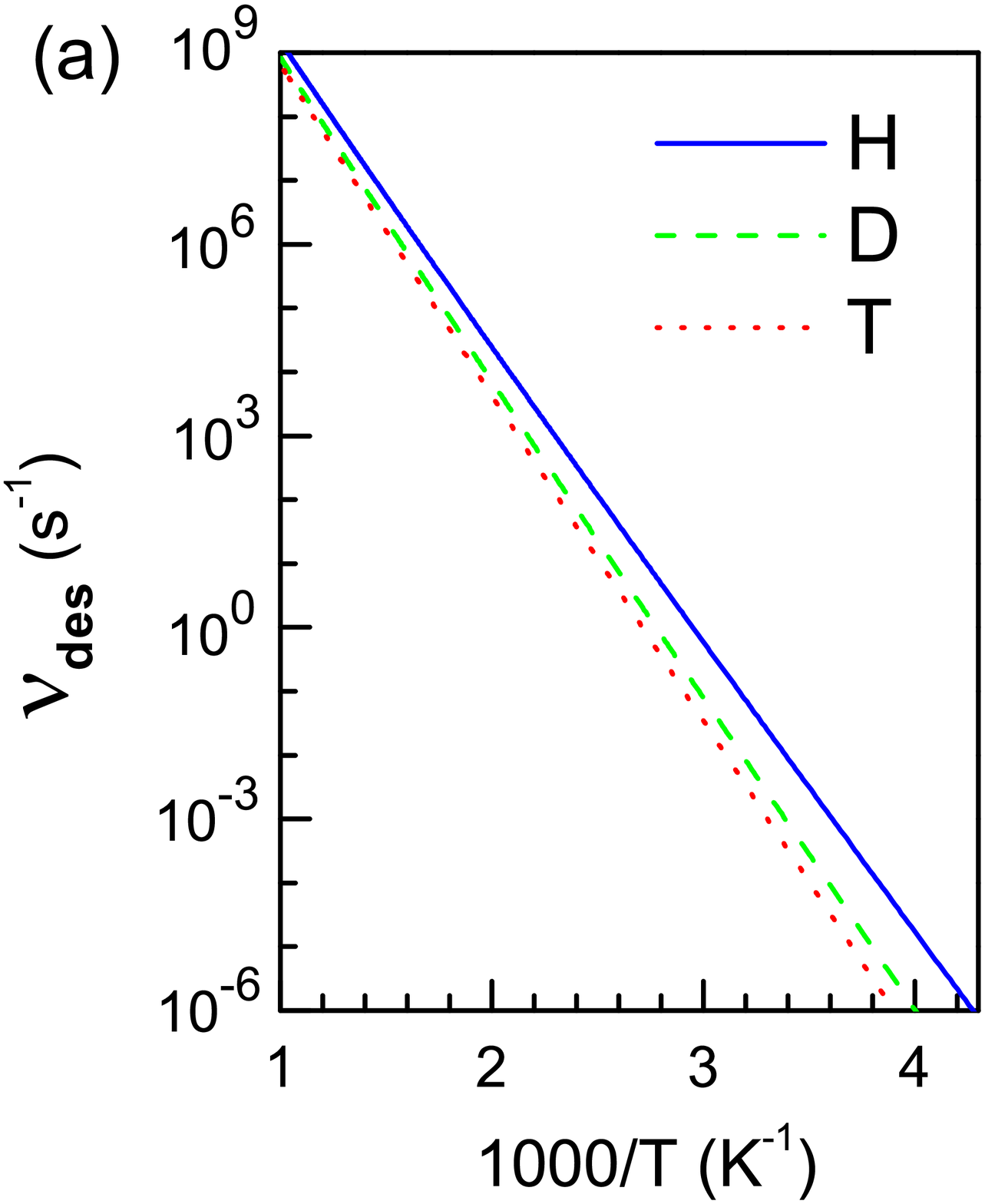}}
\scalebox{0.3}[0.3]{\includegraphics{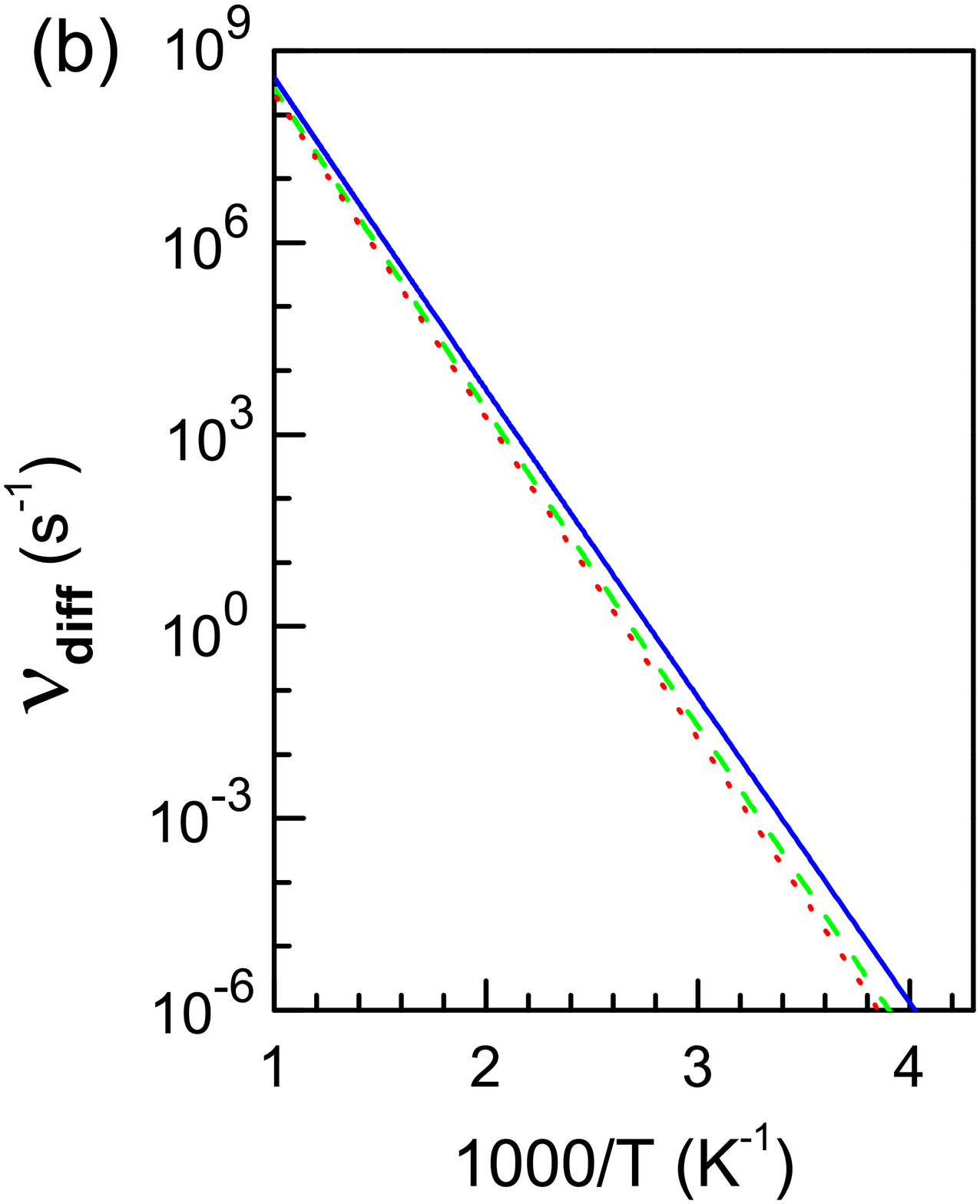}}
\caption{\label{Jumpfreq}(Color online) Variation of the (a)
desorption frequencies ($v_{des}$) and (b) diffusion frequencies
($v_{diff}$) of hydrogen-isotope monomers from the
quantum-mechanically modified hTST with respect to the inverse of
temperature.}
\end{figure}

 The desorption frequencies ($v_{des}$) and diffusion frequencies
($v_{diff}$) for the hydrogen-isotope monomers are shown in Fig.
\ref{Jumpfreq}, where the $v_{qm}^*$ is used as the prefactor. Both
$v_{des}$ and $v_{diff}$ decrease with increasing monomer mass. This
isotope effect is due to the isotope effect in the vibrational
zero-point energy correction, as shown in Tab. \ref{ZPE}. The $v$ is
exponentially dependent on the activation energy and linearly
dependent on the $v_{qm}^*$; thus, the small reversed isotope effect
in the $v_{qm}^*$s for the desorption (Fig. \ref{Prefactor}(a)) does
not compete with the normal isotope effect in the zero-point energy
corrections. For each isotope, the desorption frequencies are larger
than the corresponding diffusion frequencies because the desorption
activation energy is less than the diffusion activation energy from
Equ. \ref{activation}. Additionally, the desorption prefactor is
also larger than the diffusion prefactor. At room temperature (298
K), the $v_{diff}$s for H, D and T monomers are $1.6\times10^{-3}$,
$4.8\times10^{-4}$ and $3.0\times10^{-4}$ s$^{-1}$, respectively;
the $v_{des}$s are $1.4\times10^{-2}$, $1.4\times10^{-3}$ and
$5.5\times10^{-4}$ s$^{-1}$, respectively.

\par
\begin{figure}[htbp]
\scalebox{0.45}[0.45]{\includegraphics{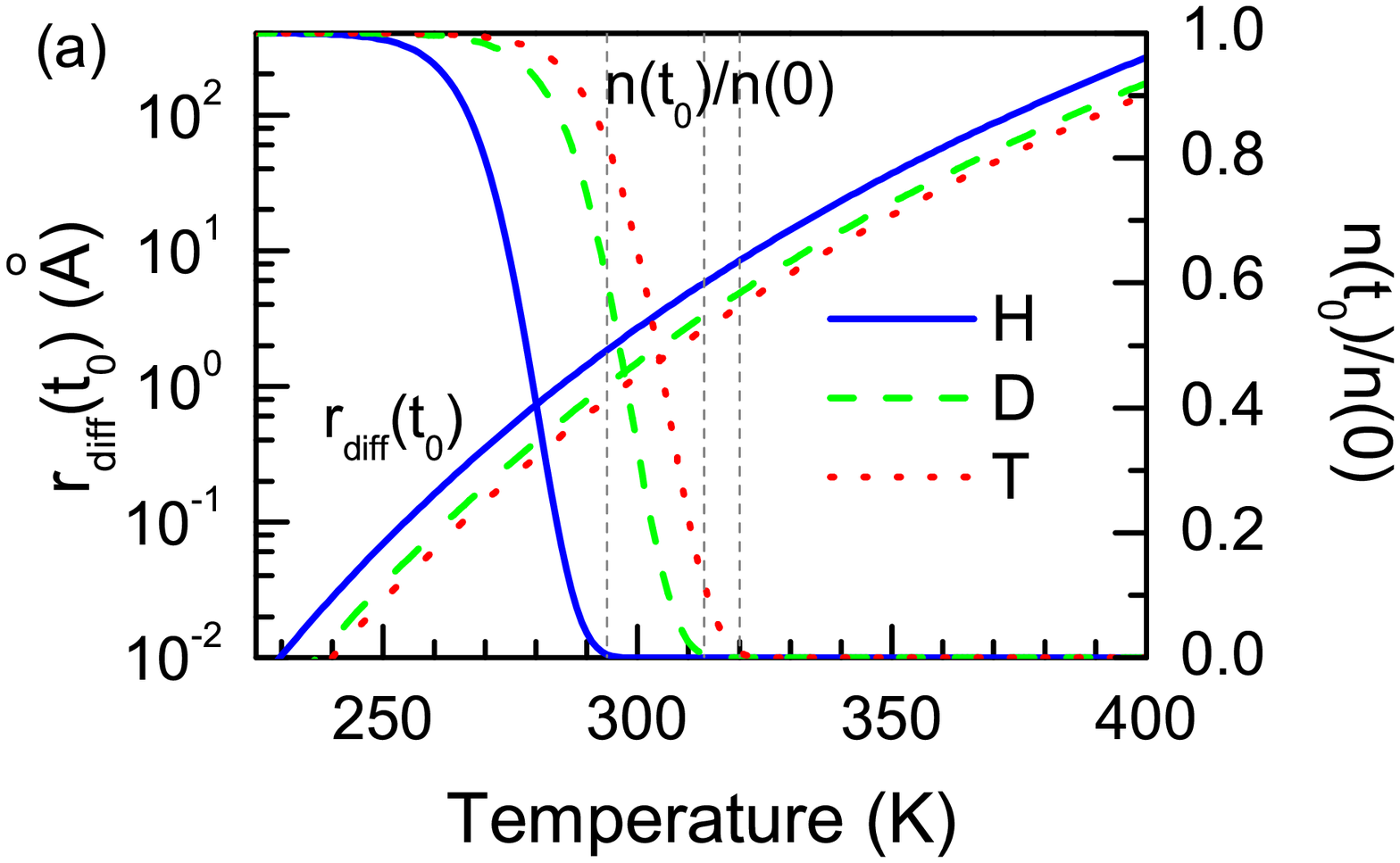}}
\scalebox{0.3}[0.3]{\includegraphics{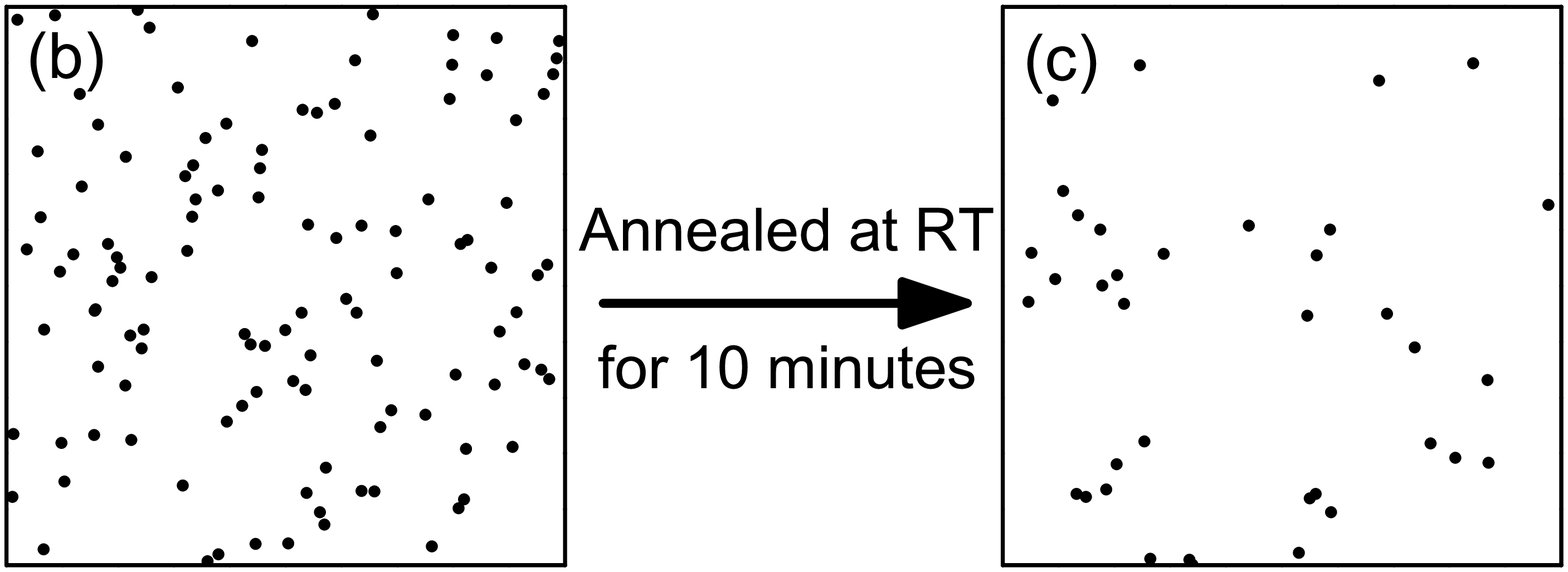}}
\caption{\label{Anneal}(Color online) (a) Variation of the diffusion
radius and the relative residual monomer number with respect to the
annealing temperature ($t_0$ = 600s). The three gray dashed vertical
lines at the complete-desorption temperatures of H, D and T monomers
(294, 312 and 320 K, respectively) are intended as a visualization
tool. (b), (c) The visualization of the annealing process in (a) for
D monomers at room temperature (298 K) for 10 minutes. The area of
the graphene surface is $100 nm \times 100 nm$, and the initial
coverage is set to be 0.03\% ($n(0) = 114$), according to the
experiment in Ref. [\onlinecite{hornekaer97}].}
\end{figure}

 After a number of hydrogen monomers are deposited onto a
graphene or graphite surface and if the system is annealed at a
constant temperature $T_0$ for a time interval $t_0$, the monomers
are desorbed from the surface or diffuse away from the initial site.
The kinetic properties of hydrogen monomers under this annealing
condition were simulated by Equ. \ref{n_anneal} and \ref{diffusion
radius}, with $t_0$ taken to be 10 minutes (600 s) in accordance
with Hornek\ae{r}'s experiment.\cite{hornekaer97} Fig.
\ref{Anneal}(a) shows the variations of the relative residual
monomer number ($n(t_0)/n(0)$) and the diffusion radius ($r_{diff}$)
with respect to $T_0$. The curves of $n(t_0)/n(0)$ and $r_{diff}$
present an isotope effect that a hydrogen monomer of lower mass is
desorbed from the surface and diffuses on the surface more easily.
During a 10-minute annealing, significant desorption does not occur
in our simulation until $T_0
>$ 250 K. However, the monomers are nearly completely desorbed
at $T_0 >$ 294 K for H, $T_0 >$ 312 K for D and  $T_0 >$ 320 K for
T. Although the $r_{diff}$(600s) increases rapidly with $T_0$, the
monomers (H, D and T) on the graphene should be diffusionally
immobile at any $T_0$ because the diffusion radius for each
hydrogen-isotope monomer is very short at temperatures below the
complete-desorption temperature, above which there will be no
residual monomer left for measurement. This type of diffusional
immobility matches experimental measurements obtained on D
monomers.\cite{hornekaer97,baouche125} The annealing process of D
monomers on graphene at 298 K is visualized in Fig. \ref{Anneal} (b)
and (c) in order to compare with Hornek{\ae}r's experiment in Ref.
[\onlinecite{hornekaer97}]. In accordance with this experiment, the
initial coverage (monomer number) is set to be 0.03\% ($n(0)=114$ in
a $100 nm \times 100 nm$ surface area) (Fig. \ref{Anneal}(b)). After
10 minutes, only 33\% of the monomers remain on graphene, a value
that is comparable to the 20\% observed in Hornek{\ae}r's
experiment. $n(t_0)/n(0)$ exponentially decreases with temperature,
and the value of 20\% corresponds to the annealing temperature of
301 K in this simulation. The discrepancy between simulation and
experiment may arise from the fact that a small fraction of the
monomers will be desorbed during the heating and cooling steps in
the annealing process; the effect of STM measurement will also
result in additional desorption of hydrogen monomers, which is not
considered in this simulation; the interlayer interaction makes the
hydrogen monomers easier to escape from graphite surface in
experiments. As mentioned in the methodology part, the interlayer
interaction in bilayer graphene will lower the potential barriers by
about 8 meV. This is because the bilayer graphene is less flexible
than the mono-layer graphene,\cite{boukhvalov21} which makes the
mono-layer graphene more reactive to bond with the hydrogen monomer.
If a correction of -8 meV is considered in the desorption activation
energy, $n(t_0)/n(0)$ equals 20\% at 299 K. Thus, the interlayer
interaction just results in temperature difference of couples of
Kelvins, which indicates its weak effect on the kinetic properties
of the hydrogen monomer and supports the validity of the mono-layer
structural model used in this report.

\par
\begin{figure}[htbp]
\scalebox{0.25}[0.25]{\includegraphics{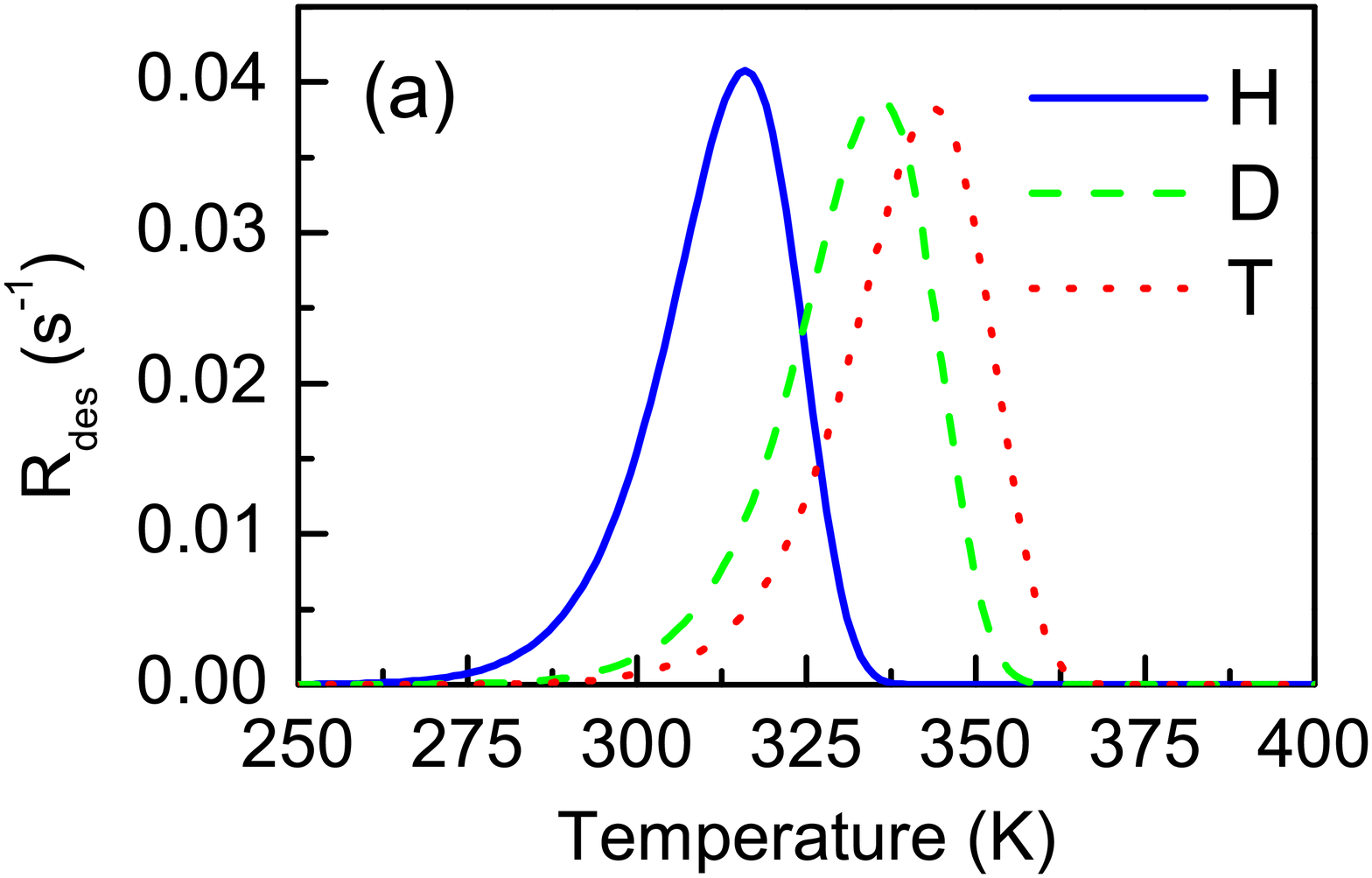}}
\scalebox{0.25}[0.25]{\includegraphics{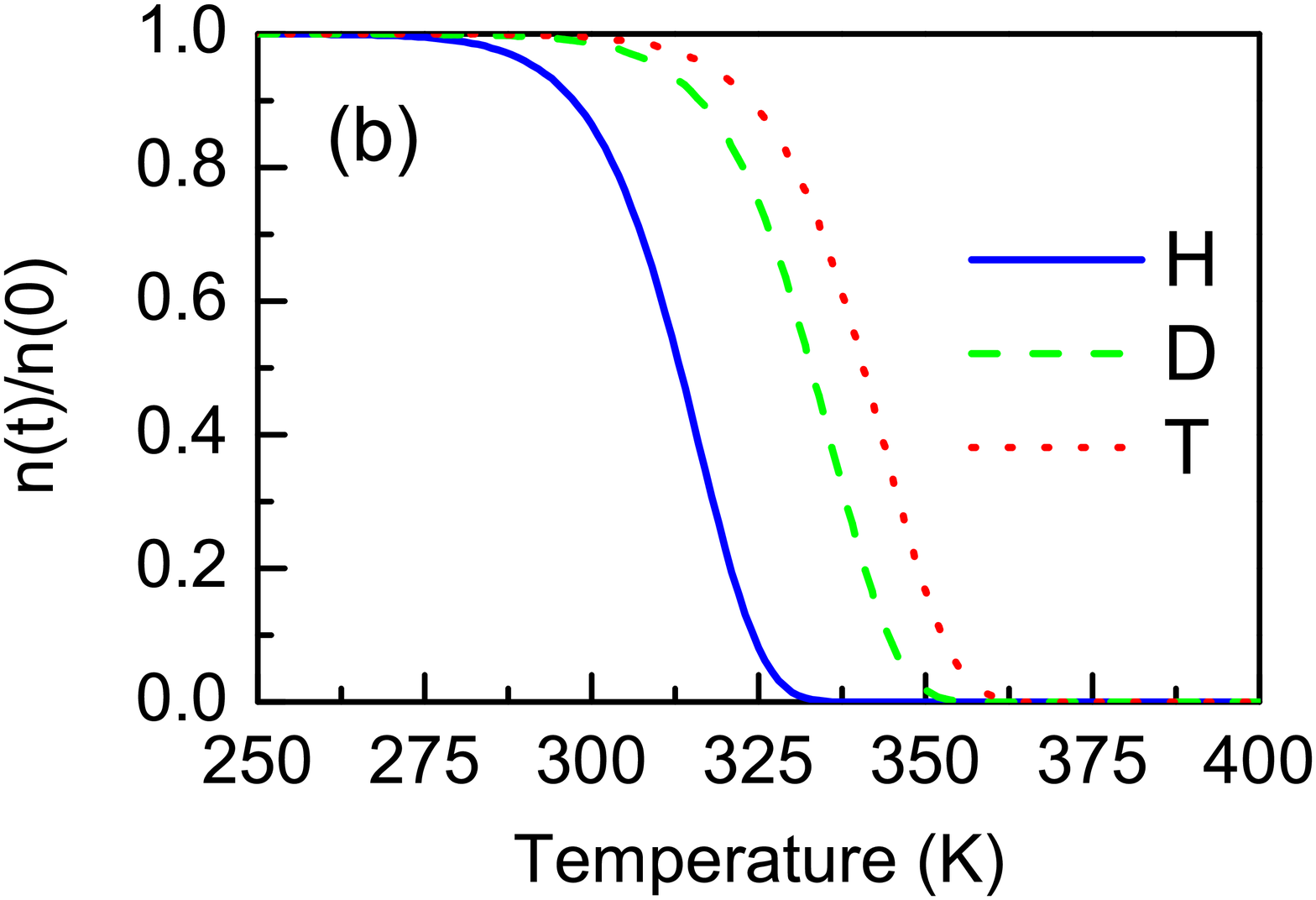}}
\caption{\label{TDS}(Color online) (a) Simulated thermal desorption
spectra for hydrogen-isotope monomers with the heating rate $\alpha$
= 1.0 K/s. (b) The residual monomer number during the thermal
desorption.}
\end{figure}

 In some experimental measurements of kinetic properties, the
temperature is increased at a constant rate ($\alpha$). The kinetic
properties of hydrogen monomers under this constant-rate heating
condition were simulated by Equ. \ref{n_heating} and \ref{desorption
rate}, with $\alpha$ taken to be 1.0 K/s. From Equ. \ref{desorption
rate}, the relative desorption rate ($R_{des}(T,t)$, Fig.
\ref{TDS}(a)) is determined both by the desorption frequency
($v_{des}(T)$) and the residual monomer number ($n(t)$). On heating,
$v_{des}(T)$ increases with temperature or elapsed time (Fig.
\ref{Jumpfreq}) and $n(t)$ decreases (Fig. \ref{TDS}(b)); thus,
there will be a desorption peak for each hydrogen-isotope monomer in
the desorption spectrum of $R_{des}$. The desorption peak for the H
monomer is at 316 K, for the D monomer at 336 K and for the T
monomer at 344 K. In the desorption spectra, an isotope effect can
be seen that a hydrogen monomer of lower mass is desorbed from the
surface more easily, which is the same as the isotope effect in the
annealing process. The same as discussed in the previous paragraph,
the weak interlayer interaction will only make the desorption
spectra move to low temperatures by about 2 K.

\par In addition, when the coverage of hydrogen atoms on graphene increases,
hydrogen dimers will form and dominate on the
surface.\cite{hornekaer96,zecho42,zecho117} In the desorption
spectra under constant-rate heating (1.0 K/s), there are two peaks
at 445 (490) K and 560 (580) K for H (D) dimers, which is totally
different from the desorption spectra of H (D) monomer in Fig.
\ref{TDS}. The kinetic properties of hydrogen dimers on graphene and
the effect of dimer-dimer interaction have been investigated by us
using nearly the same method, which will appear in a forthcoming
paper. Generally, the isotope effects, localization nature of the
vibrations of hydrogen dimers and the dependence of the kinetic
properties on the localized modes are the same as those of the
hydrogen monomer. This means the investigation on the kinetic
properties of hydrogen monomer on graphene is instructive to other
more complex situations.

\section{CONCLUSIONS}
The kinetic properties of hydrogen-isotope (H, D and T) monomers on
graphene were simulated using a composite {\itshape{ab initio}}
method consisting of density functional theory, density functional
perturbation theory and harmonic transition state theory. The
simulations were based on the potential barriers calculated from DFT
and on the phonon spectra calculated from DFPT, which were used
together to analyze calculated vibrational zero-point energy
corrections, exponential prefactors (quantum-mechanically modified
and classical-limit), jump frequencies and simulated kinetic
properties. The chemisorption state and desorption saddle-point
state both have a magnetic moment of 1.0 $\muup_B$, while the
diffusion saddle-point state is nonmagnetic. From the electronic DOS
analysis, in the desorption process, the hybridization of the
hydrogen monomer with graphene is reduced and the contribution of
the H(1$s$) orbital to the total magnetic moment increases. And in
the diffusion process, when leaving from the chemisorption state to
the diffusion saddle-point state, this hybridization also decreases;
the transferred electrons from H to graphene become more and more
itinerant; the total magnetic moment decreases from 1.0 $\muup_B$
down to 0.0 $\muup_B$. The spectra of the localized vibrational
modes of hydrogen are hydrogen-mass-dependent, which results in the
isotope effects in the kinetic properties (prefactor, activation
energy, jump frequency, desorption rate and diffusion radius) of H,
D and T monomers. The zero-point energy correction decreases with
increasing monomer mass and is larger for desorption than for
diffusion. This results in the isotope effect that a hydrogen
monomer of lower mass is desorbed and diffuses more easily (with
lower activation energies), and in that the desorption frequency of
each hydrogen-isotope monomer is larger than its diffusion
frequency. In the simulated 10-minute annealing of hydrogen-isotope
monomers on graphene, the monomers are quite diffusionally immobile
at temperatures lower than the complete-desorption temperature, and
a large fraction (67\%) of the D monomers on graphene will be
desorbed at the annealing temperature of 298 K. These theoretical
results are very close to those obtained by experimental
observations of hydrogen adatoms on graphite
surfaces.\cite{hornekaer97,baouche125} The thermal desorption
spectra ($\alpha=1.0$ K/s) of the hydrogen-isotope monomers have
also been simulated. The resulting desorption peak for H is at 316
K, for D is at 336 K and for T is at 344 K. The precise simulation
of the kinetic properties of the hydrogen-isotope monomers on
graphene described in this report can help to further precise
studies of more complex hydrogenation processes. The precise
theoretical predictions for radioactive tritium adatoms are
especially useful because it is dangerous to conduct experiments on
them in conventional laboratories.

\begin{acknowledgments}
The first author (Huang) wishes to thank Liv Horkek{\ae}r and K.
Toyoura for helpful email exchanges. This work was supported by the
special Funds for Major State Basic Research Project of China (973)
under grant No. 2007CB925004, 863 Project, Knowledge Innovation
Program of Chinese Academy of Sciences, and by Director Grants of
CASHIPS. Part of the calculations were performed at the Center of
Computational Science of CASHIPS and at the Shanghai Supercomputer
Center.
\end{acknowledgments}

\bibliography{basename of .bib file}

\begin{table}[htbp]
\caption{\label{ZPE}Vibrational zero-point energy corrections
(${\Delta}F_{vib}(0)$) in the activation energies (in eV) for the
desorption (des.) and diffusion (dif.) of hydrogen-isotope (H, D and
T) monomers.}
\begin{ruledtabular}
\begin{tabular}{cccc}
  & ${\Delta}F^H_{vib}(0)$ & ${\Delta}F^D_{vib}(0)$& ${\Delta}F^T_{vib}(0)$\\
\hline
des. & 0.205 & 0.140 & 0.113  \\
dif. & 0.114 & 0.084 & 0.072  \\
\end{tabular}
\end{ruledtabular}
\end{table}

\end{document}